\documentclass[a4paper,11pt]{article}
\usepackage{pos,wrapfig,bm}

\title{CT25: the new generation of CTEQ-TEA PDFs}
 \ShortTitle{A. Ablat et al., CT25: the new generation of precision CTEQ-TEA PDFs}

 \author[a]{Alim Ablat}
 \author[b]{Daniel Adamiak}
\author[c,*]{Aurore Courtoy\footnotetext[1]{Presenter and corresponding author}}
\author[b,d]{Sayipjamal Dulat}
\author[b]{Yao Fu}
\author[e,*]{Marco Guzzi}
\author[f]{T.~J.~Hobbs}
\author[b]{Joey Huston}
\author[b]{Maximiliano Ponce Chavez}
\author[b]{Kirtimaan Mohan}
\author[b,*]{Pavel Nadolsky}
\author[b,*]{Tanishq Sharma}
\author[b]{Dan Stump}
\author[g]{Keping~Xie}
\author[b]{C.-P. Yuan}

\affiliation[a]{School of Physics and Electrical Engineering, Kashi University,
Kashi, Xinjiang 844000, China}

\affiliation[b]{Department of Physics and Astronomy, Michigan State University,
East Lansing, MI 48824, USA}

\affiliation[c]{Instituto de F\'isica, Universidad Nacional Aut\'onoma de M\'exico,\\
Apartado Postal 20-364, Ciudad de M\'exico 01000, Mexico}

\affiliation[d]{School of Physics Science and Technology, Xinjiang University,
Urumqi, Xinjiang 830046, China}

\affiliation[e]{Department of Physics, Kennesaw State University, 
Kennesaw, GA 30144, USA}

\affiliation[f]{High Energy Physics Division, Argonne National Laboratory,
Lemont, IL 60439, USA}

\affiliation[g]{Tsung-Dao Lee Institute \& School of Physics and Astronomy, Shanghai Jiao Tong University,\\ Shanghai 201210, China\looseness=-1}

\emailAdd{aurore@fisica.unam.mx}
\emailAdd{mguzzi@kennesaw.edu}
\emailAdd{nadolsky@msu.edu}
\emailAdd{shar1157@msu.edu}

\abstract{This contribution reviews CT25, a new generation of CTEQ--TEA parton distribution functions (PDFs) for precision QCD. With its first member sets released in 2025, the CT25 family addresses common uses of parton distributions in diverse contexts. We describe the CTEQ--TEA Parton Cloud, an online database of exploratory global fits to precision QCD measurements, from which we select PDF solutions for general- and special-purpose calculations. These solutions are published as complementary error PDF ensembles---the baseline CT25, CT25m, CT25pd, and ensembles with conservative priors---optimized for distinct uses. They implement new approaches to quantifying uncertainties and different treatments of theoretical, parametrization, and nuclear-target uncertainties. In particular, they provide complementary frameworks for exploring PDF uncertainties at large partonic momentum fractions relevant to HL-LHC and EIC studies. Further details will appear in the forthcoming CT25 publication.}

\FullConference{The 33rd International Workshop on Deep Inelastic Scattering and Related Subjects (DIS2026)\\
4 - 8 May 2026\\
Bologna, Italy\\}

\begin{document}
\maketitle

\begin{figure}[tb]
    \centering
\includegraphics[width=0.8\linewidth]{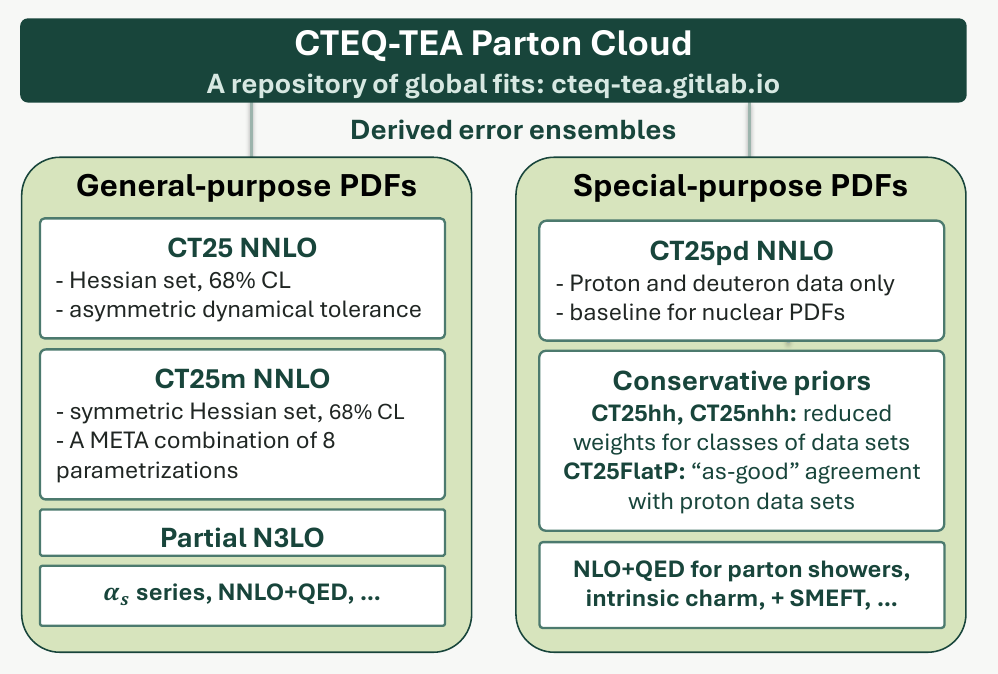}
    \caption{The CTEQ-TEA Parton Cloud is a repository of a wide variety of PDF fits, from which families of general- and special-purpose PDF sets are derived. Graphic generated with the aid of GPT 5.6 Sol.\label{fig:CTPartonCloud}}
\end{figure}

Precision measurements planned for the high-luminosity phase of the LHC raise the standards for parton distribution functions (PDFs) used in experimental analyses. In preparation for the HL-LHC, the CTEQ-Tung et al. (CTEQ-TEA) group \cite{CTwebsite} is releasing a new generation of nucleon PDFs named CT25. The first default error ensemble, CT25 NNLO, was released in 2025 \cite{Ablat:2025gdb}, following several years of exploratory studies \cite{Sitiwaldi:2023jjp,Ablat:2023tiy,Ablat:2024uvg,Ablat:2025gbp}.
All CT25 PDFs are now published on the CTEQ-TEA website \cite{CTwebsite} and will be submitted to the LHAPDF library.
In this contribution we summarize the key motivations and results of the CT25 analysis, which are further detailed in a journal publication~\cite{ct25nnlo}.

The CT25 study emphasizes a comprehensive exploration of PDFs at NNLO accuracy. In parallel, we develop PDFs at partial N$^3$LO accuracy; see a complementary contribution~\cite{disN3LO}. We choose this strategy because many factors already present at NNLO have effects comparable to those of the partly known N$^3$LO QCD contributions. To gauge the PDF uncertainty fully, these factors must be explored exhaustively. This view motivates the CTEQ-TEA Parton Cloud, a new comprehensive database that provides public PDF solutions obtained with a variety of parametrizations, data selections, and methodological assumptions.

Figure~\ref{fig:CTPartonCloud} illustrates the CTEQ-TEA Parton Cloud. It contains a large number (currently about 400) of best-fit PDF solutions compatible with the fitted experimental data sets, obtained by varying the data selection, PDF functional forms, and theoretical priors such as positivity. A representative sampling of those variables~\cite{Courtoy:2022ocu} is hence achieved through the Parton Cloud. A user can explore all relevant solutions explicitly to estimate the underlying probability distribution while accounting for the descriptive information that is also provided. In addition to releasing the Cloud best-fit solutions, we publish more traditional general-purpose ensembles such as CT25 NNLO, obtained by fitting several fixed parametrization forms selected from the Cloud and applying 
established tolerance definitions, defined below. Finally, we publish special-purpose ensembles that complement the general-purpose ones, such as NNLO QCD PDFs with NLO QED contributions. Among these, PDFs with conservative priors can be used to estimate the sensitivity of results to missing information in global fits.

Altogether, the CT25 family moves beyond reliance on a single tolerance prescription to represent PDF uncertainty across all flavors and partonic momentum fractions. Our upcoming publication \cite{ct25nnlo} contains detailed recommendations for choosing concrete CT25 ensembles for a typical problem at hand.
The default CT25 NNLO ensemble is suitable for most situations, while other ensembles provide complementary insights as needed. Uncertainties associated with the choice of parametrization become increasingly important in regions where data provide limited direct constraints. A direct consequence is that the extended CT25 family targets improved predictions and uncertainty definition in the large-$x$ region for HL-LHC and EIC scattering processes.

This contribution summarizes three CTEQ-TEA presentations at DIS 2026, starting with a review of the default CT25 NNLO analysis and moving to discussions of the Parton Cloud, tolerance, and PDF+$\alpha_s$ analysis. We also review other new error ensembles: CT25m, which probes PDF parametrization dependence; CT25pd, based only on a proton-deuteron data set; CT25FlatP, which collects "as-good" fits to proton data sets in our baseline; and, finally, CT25hh and CT25nhh fits with modified statistical weights for hadron-hadron and complementary data sets.

\vspace{-6pt}\section{The nominal CT25 NNLO ensemble \label{sec:CT25NNLO}}

We start by reviewing the default CT25 analysis, which is the successor of the CT18 family \cite{Hou:2019efy} and is closest to CT18 in its methodology. The CT25 parametrization is selected from about 300 exploratory fits obtained with varied PDF forms, data selections, and methodological assumptions. Within this collection, CT25 uses a choice of PDF parametrizations that yields one of the lowest log-likelihood ($\chi^2$) values [the orange star in the left panel of Fig.~\ref{fig:alphas_scan_290forms}]. An even lower $\chi^2$ can be achieved by relaxing either positivity or the requirement of a common power law for the large-$x$ falloff of the $u$ and $d$ PDFs, both imposed in CT25 for theoretical reasons.

The CT25 fit extends the dataset baseline of the CT18 analysis with new Drell--Yan, $t\bar t$, and inclusive-jet measurements (13 precision LHC data sets at center-of-mass energies of 5.02, 8, and 13~TeV selected based on \cite{Sitiwaldi:2023jjp,Ablat:2023tiy,Ablat:2024uvg,Ablat:2025gbp}),  together with the SeaQuest fixed-target Drell--Yan measurement, and removes four charged-current DIS data sets on iron targets. The overall agreement with the extended baseline is reasonable but not ideal: the most precise new experiments often have somewhat elevated $\chi^2/N_{\rm pt}$, continuing a pattern already seen in CT18. The new measurements increase $\chi^2/N_{\rm pt}$ for several legacy data sets, including CDHSW, NuTeV, and E866 deuteron-to-proton Drell--Yan data, while improving the description of E866 proton--proton Drell--Yan, the D0 lepton asymmetry, the CDF $Z$ rapidity distribution, and several LHC weak-boson and asymmetry measurements.

\begin{figure}[tb]
    \centering
\includegraphics[height=160pt]{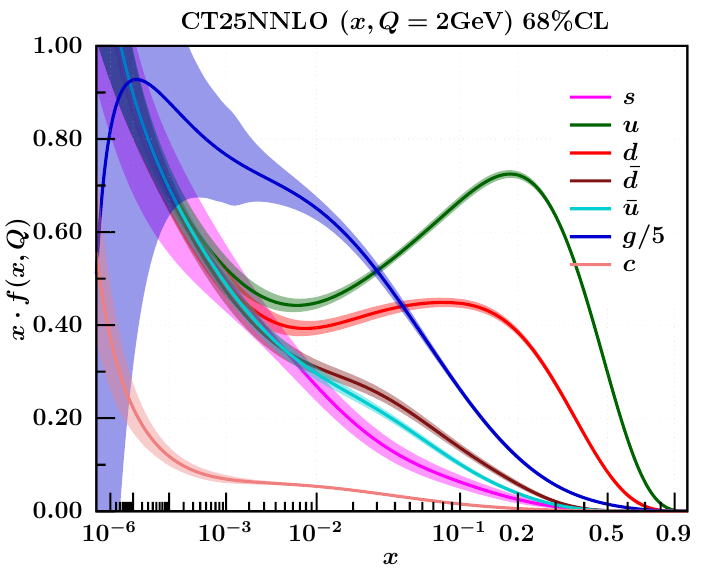}
\includegraphics[height=160pt]{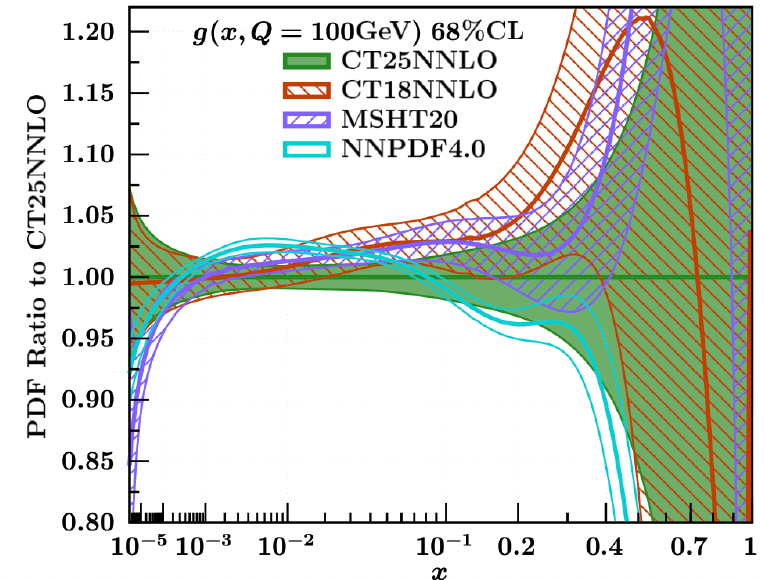}
\caption{
Left: CT25 NNLO parton distribution functions at $Q=2$ GeV. Right: The CT25 NNLO gluon at $Q=100$ GeV compared with CT18 (red), MSHT20 (purple), and NNPDF4.0 (cyan), shown as ratios to CT25 NNLO.
}
\label{fig:CT25NNLO-PDFs}
\end{figure}

The left panel of Fig.~\ref{fig:CT25NNLO-PDFs} shows the CT25 NNLO PDFs at $Q=2$ GeV. The right panel compares its gluon PDF at 100 GeV with other recent ensembles. Other comparisons are included in Nadolsky's slides. In general, the CT25 central PDFs and uncertainty bands remain consistent with CT18, while its uncertainties are smaller because of new data sets and the modified tolerance prescription discussed below. CT25 is closer to CT18Z than to nominal CT18. Adding LHC inclusive jet measurements and removing the inclusive CCFR and CDHSW neutrino--nucleus DIS data favor a softer gluon for $x\gtrsim10^{-2}$. The new LHC Drell--Yan measurements increase the strange PDF at $x\gtrsim0.1$ relative to CT18.

Figure~\ref{fig:CT25NNLO-PDFs} illustrates several trends summarized as follows. Both the CT25 PDFs and parton luminosities (not displayed) are broadly consistent with CT18~\cite{Hou:2019efy}, MSHT20~\cite{Bailey:2020ooq}, and NNPDF4.0~\cite{NNPDF:2021njg}, but differences relevant to precision LHC phenomenology remain visible. In the slides, we consider two metrics for the normalized distance between PDF sets, respectively akin to the cumulative $L_1$ distance and the uncertainty-normalized cumulative $L_1$ distance. These distances show that, at $Q=2$~GeV, CT18 remains closest to CT25 for the up- and down-quark combinations. MSHT20 is closest for the strange sea and gluon in this metric, while NNPDF4.0 is generally the most distant. Relative to CT18, CT25 moves closer to NNPDF4.0 and MSHT20 for the down and strange distributions, closer to NNPDF4.0 for the gluon, while the up distribution generally moves farther away, and the gluon moves farther from MSHT20.

The CT25 luminosity uncertainties are comparable to those of MSHT20. New LHC data in the CT25 NNLO baseline reduce the gluon-gluon luminosity in the Higgs region by about 2.5\%, comparable to the additional 1.2\% reduction from partial N$^3$LO DIS contributions \cite{Guzzi:2026ktp}. The total reduction of the $gg$ luminosity at $x\sim10^{-2}$ and $Q\approx100$ GeV in our preliminary fits is therefore about 4\%.

At $\sqrt{s}=14$~TeV, predictions for $Z$, $W^+$, $W^-$, $t\bar t$, and gluon-fusion Higgs production populate distinguishable correlated regions for CT18, CT18Z, CT25, MSHT20, and NNPDF4.0. See slides.

\vspace{-6pt}\section{Tolerance, convexity, systematic uncertainties}

The key advance of the CT25 methodology concerns its definition of PDF uncertainty, or tolerance. In CT18 and earlier analyses, PDF errors were estimated using a two-tier tolerance that accounts for agreement with both the global data set and individual subsets \cite{Lai:2010vv}. Recent developments, including the CT25 PDF+$\alpha_s$ study \cite{Ablat:2025gbp}, suggest a more nuanced approach for future PDF applications.

\begin{figure}[tb]
    \centering
    \includegraphics[height=130pt]{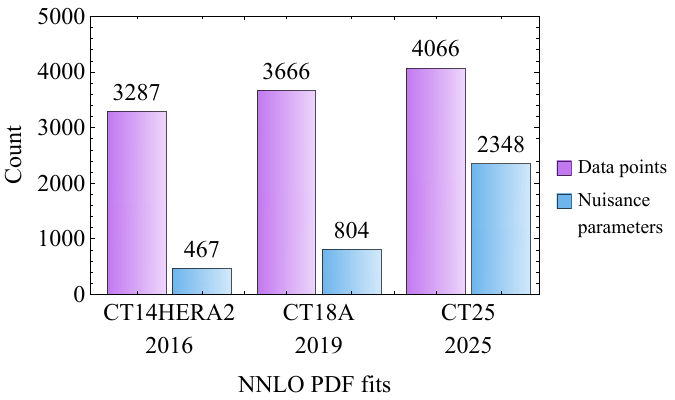}
    \includegraphics[height=130pt]{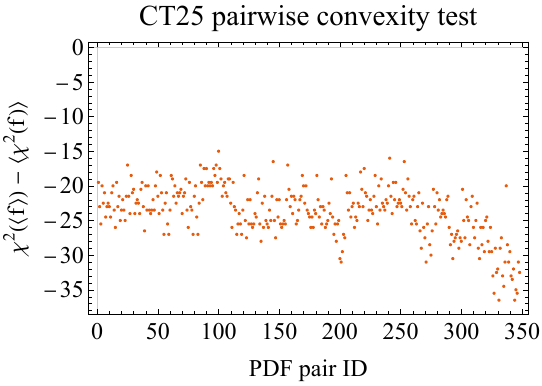}\quad
    \caption{Left: 
    Numbers of data points and nuisance parameters in the CT14HERA2, CT18A, and CT25 NNLO fits.
    Right: $\chi^2$ convexity test for PDFs in the CTEQ-TEA Parton Cloud. The log-likelihood $\chi^2(\langle f \rangle)$ of an average of two random members of the CTEQ-TEA Parton Cloud tends to be lower than the average $\langle \chi^2(f)\rangle $ of log-likelihoods for these members. 
     \label{fig:NuisanceParsConvexity}} 
\end{figure}

At the HL-LHC, a typical inference problem will determine Standard Model (SM) or Beyond the Standard Model  (BSM) parameters of interest (POIs), represented by a vector $\bm{\theta}$, together with nucleon-structure parameters $\bm{\pi}$ and parameters $\bm{\lambda}$ quantifying experimental and theoretical systematic effects. The problem generally requires a global pre-LHC+LHC data set $D$ and finding the joint posterior probability from Bayes' theorem:
\begin{equation}
P(\bm{\theta},\bm{\pi},\bm{\lambda} \mid D)
=
\frac{P(D \mid \bm{\theta},\bm{\pi},\bm{\lambda})\,
P(\bm{\theta},\bm{\pi},\bm{\lambda})}{P(D)}.
\label{fit0}
\end{equation}
Because such a \textit{full} Bayesian determination has not been computationally feasible, two \textit{practical} Bayesian approaches are commonly pursued:
\begin{enumerate}
  \item A PDF fit to the global data set $D$ determines the optimal PDF model $\bm{\pi}_0$ and POIs $\bm{\theta}_0$ as a maximum a posteriori (MAP) solution using a surrogate, typically linear, model $\widetilde{\bm{\lambda}}$ for systematic factors:
        \begin{align}
          P(\bm{\theta},\bm{\pi},\widetilde{\bm{\lambda}}\mid D)
          &\propto
          P(D\mid\bm{\theta},\bm{\pi},\widetilde{\bm{\lambda}})
          P(\bm{\theta})
          P(\bm{\pi})\,
          \widetilde P(\widetilde{\bm{\lambda}}),
          &
          \{\bm{\pi}_0, \bm{\theta}_0\}
          &=\operatorname*{arg\,max}_{\bm{\theta}, \bm{\pi},\widetilde{\bm{\lambda}}}
          P(\bm{\theta},\bm{\pi},\widetilde{\bm{\lambda}}\mid D).
          \label{fit1}
        \end{align}
       The CT25 global PDF analysis follows this approach. Here, the surrogate model $\widetilde{\bm \lambda}$ refers to large correlation matrices estimating experimental systematic uncertainties. The associated nuisance parameters ($N_\nu=2348$) number more than half the measurements ($N_\textrm{pt}=4066$); see the left panel of Fig.~\ref{fig:NuisanceParsConvexity}.
  \item A complementary (not exclusive) approach is a POI estimate from a \textit{single} LHC data set $\overline D$ using a surrogate  PDF ensemble $\widetilde{\bm{\pi}}$ approximating PDF uncertainties: 
        \begin{align}
        P(\bm{\theta},\bm{\lambda},\widetilde{\bm{\pi}}\mid \overline D)
          &\propto
          P(\overline D\mid\bm{\theta},\bm{\lambda},\widetilde{\bm{\pi}})
          P(\bm{\theta})
          P(\bm{\lambda})\,
          \widetilde P(\widetilde{\bm{\pi}}),
          &
          \bm{\theta}_0
          &=\operatorname*{arg\,max}_{\bm{\theta}, \widetilde{\bm{\pi}},\bm{\lambda}}
          P(\bm{\theta},\bm{\lambda},\widetilde{\bm{\pi}}\mid \overline D).
          \label{fit2}
        \end{align}
\end{enumerate}
Traditionally, the surrogate PDF model $\widetilde{\bm{\pi}}$ is provided as either Hessian or MC error PDFs. However, the CTEQ-TEA Parton Cloud includes many fits with excellent $\chi^2$, and linear combinations of Cloud members tend to provide equally good fits because of the log-likelihood convexity observed in the preliminary CT25 phase [right panel of Fig.~\ref{fig:NuisanceParsConvexity}]. Traditional PDF tolerances do not account for this variety. Experimental analyses can instead profile the POIs with $\Delta\chi^2=1$ if the "PDF prior" $\widetilde{\bm{\pi}}$ represents the full span of Cloud solutions.

On the downside, the "practical Bayes" methods (\ref{fit1}) and (\ref{fit2}) reduce the "full Bayes" problem (\ref{fit0}) by using linear approximations for either $\bm{\lambda}$ or $\bm{\pi}$ and neglecting their cross-correlations. When the numbers of POIs ($N_\textrm{poi}$), nuisance parameters ($N_\nu$), and associated priors ($N_\textrm{prior}$) are comparable to the number of data ($N_\textrm{pt}$), the expected log-likelihoods for fitted and predicted samples differ nontrivially from $\langle \chi^2\rangle=N_\textrm{pt}$, valid as $N_\textrm{pt}/(N_\textrm{poi}+N_\nu)\to\infty$ \cite{Adamiak:2026chi2}:
\begin{align}
\langle \chi^2 \rangle_\textrm{fitted} = N_\textrm{pt} + N_\textrm{prior} - N_\textrm{poi} - N_{\nu}; \quad 
\langle \chi^2 \rangle_\textrm{predicted} = N_\textrm{pt} + N_\textrm{prior} + N_\textrm{poi} + N_{\nu}.
\end{align}
These formulas show deviations from asymptotic goodness-of-fit estimators when an analysis includes thousands of nuisance parameters (NPs) of experimental or theoretical origin. In the actual fit, many NPs appear idle: their posterior deviations from zero are much smaller than expected from their unit-Gaussian priors. Nevertheless, the numbers of NPs and priors are now comparable to the number of measurements, raising the stakes for understanding the underlying uncertainty.

\section{CT25 $\alpha_s$ uncertainty and clustering dependence of the dynamic tolerance}
The PDF uncertainty must also account for  the dependence on PDF functional forms and imperfect agreement (tensions) among individual data sets. Together, these factors have long justified increasing the PDF tolerance from the textbook criterion $\Delta\chi^2=1$ at 68\% probability. In preparation for the full CT25 release, Ref.~\cite{Ablat:2025gdb} examined this issue through a determination of the QCD coupling $\alpha_s$ from the CT25 baseline.

\begin{figure}[tb]
    \centering
    \includegraphics[height=155pt]{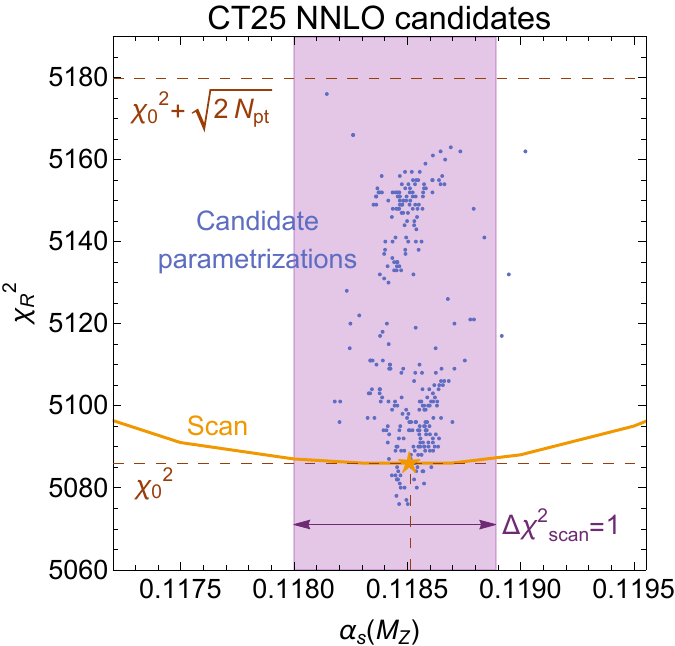}
    \includegraphics[height=155pt]{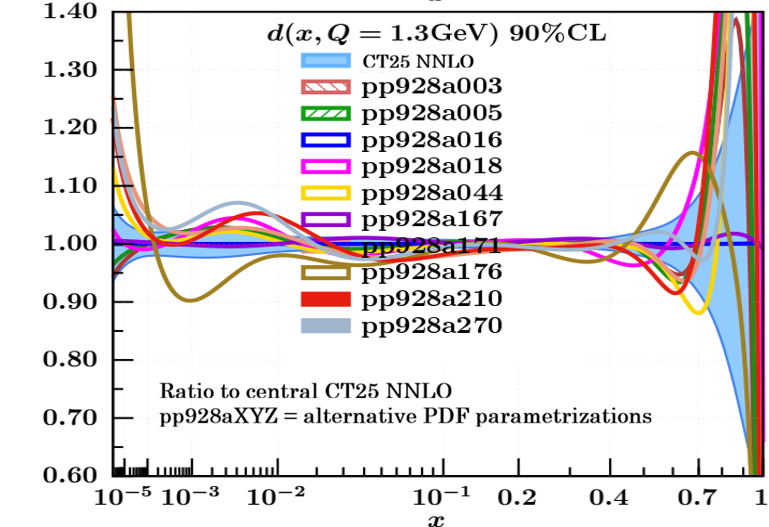}
    \caption{
    Left: Nearly all of 287 alternative cloud parametrizations explored in the CT25 NNLO analysis prefer $\alpha_s(M_Z)$ within the $\Delta \chi^2=1$ interval from the CT25 NNLO default fit. 
    $\chi^2_R$ is the $\chi^2$ component without prior penalties. 
    Right: PDFs for many parametrizations themselves fall outside of the $\Delta\chi^2=1$ error band for the CT25 NNLO fit, as demonstrated on the example of the $d$ quark PDF.   
    \label{fig:alphas_scan_290forms}}
\end{figure}

Figure~\ref{fig:alphas_scan_290forms} and the CT25 $\alpha_s$ estimates below illustrate the nuanced picture of PDF tolerance that emerges from this study. The 68\% credibility interval based on $\Delta\chi^2=1$ for the full CT25 baseline, $\alpha_s(M_Z)=0.1183(5)$, uses a single PDF parametrization and accounts only for statistical uncertainty. It is narrower than the 2023 Particle Data Group (PDG) average, $\alpha_s(M_Z^2)=0.1180(9)$. Interestingly, the $\alpha_s$ value depends little on the PDF parametrization, which we explored using 287 diverse functional forms; see the left panel of Fig.~\ref{fig:alphas_scan_290forms}. The applicability of the $\Delta\chi^2=1$ criterion is observable-specific and may fail when there is a strong correlation with PDF shapes. For example, the right panel shows that, for the PDFs themselves, the $\Delta\chi^2=1$ error with a fixed parametrization underestimates the span obtained with alternative functional forms.

Including the PDF parametrization uncertainty mildly increases $\delta\alpha_s(M_Z)$ to about $0.0007$. For real global data, a far more important error arises from disagreement among the $\alpha_s(M_Z)$ values preferred by different classes of experiments. The partial $\alpha_s$ credibility intervals at 68\% within a single fit are
$0.1151(6)$ for DIS, $0.1186(7)$ for vector-boson production, and $0.1207(5)$ for the combined sample of single-inclusive jet and $t\bar t$ production measurements. Their naive combination yields $\alpha_s(M_Z)=0.1184(4)$, resulting in an error smaller than the PDG value as well as a low $p$ value. Furthermore, this value is generally sensitive to normalizations of experimental systematic errors, see Ref.~\cite{Ablat:2025gdb}, 

These observations invite a corrective approach that accounts for tensions among the data sets. The PDG prescription for combining inconsistent measurements (see Sec.~IV.2 of Ref.~\cite{Ablat:2025gbp}) would achieve acceptable agreement at 68\% CL by inflating all errors by a factor $e_\textrm{PDG}=\sqrt{\chi^2/{\rm d.o.f.}}=\sqrt{66/3}=4.7$. As a more informed procedure, we examined a combination based on four alternative methods: dynamic tolerances based on the combined and 52 individual data sets, a Bayesian hierarchical model, and a Gaussian-mixture model. We obtained
\begin{equation}
\alpha_s(M_Z)=0.1183^{+0.0023}_{-0.0020}\qquad (68\%\ {\rm CL}),
\label{alsMZ}
\end{equation}
in good agreement with all data sets. Although this result assumes no a priori tolerance, it is consistent with a global $\Delta\chi^2\simeq20$--35 in a single-parametrization $\alpha_s$ scan.

From this study, we conclude that for real global data, data-selection choices and systematic-error models dominate, $\Delta\chi^2=1$ fails, and the 68\% CL uncertainty is closer to $\delta\alpha_s\simeq0.002$. A second significant conclusion, which extends to the entire PDF analysis, is that statistical coverage, or tolerance, requires separate conventions depending on the application or observable because of the factors highlighted in this and preceding  sections.

Among the considered possibilities, the dynamic tolerance (DT) computed from 52 published datasets provided uncertainty estimates consistent with the two Bayesian methods. 
Dynamic tolerance \cite{Martin:2009iq,Lai:2010vv,Kovarik:2019xvh} determines Hessian PDF errors by requiring acceptable quantiles of $\chi_i^2/N_{{\rm pt},i}$ for individual subsets and allows the effective $T^2$ to vary with the eigenvector direction. Depending on the granularity of the subsets, which may be constructed by mixing data points from different experiments, the DT uncertainty lies between the errors obtained with $\Delta\chi^2=1$ and the global tolerance based on the total baseline data \cite{Ablat:2025gbp}. If DT is computed with the nominal subsets published by individual experiments, it is straightforward to implement and offers a balanced uncertainty estimate, although the estimate may change under alternative data clustering and is most sensitive to the least-compatible experiments.
 
\vspace{-6pt}\section{Derived PDF ensembles of the CT25 family \label{sec:Derived ensembles}}
The preceding discussion outlines a way to construct practical error PDF ensembles from the growing number of high-quality central fits in the CTEQ-TEA Parton Cloud. This repository stores both tables of the resulting PDFs and descriptive metadata about $\chi^2$ and methodological assumptions, facilitating exploration of acceptable PDF solutions. Currently, the Cloud contains roughly 400 fits that primarily use Bernstein-polynomial parametrizations, as in CT18, with the possibility of implementing B\'ezier curves~\cite{Kotz:2025une} and other universal approximators in the near future~\cite{ct25alt}.

In the CT25 release, we construct several error ensembles from the fits with the lowest $\chi^2$ values in the sample and with diverse functional behaviors. 
\begin{enumerate}
    \item \textbf{CT25 NNLO} is a general-purpose Hessian ensemble with one PDF parametrization, obtained from comprehensive data sets with 68\% CL uncertainties defined by asymmetric dynamic tolerance.
    \item \textbf{CT25m NNLO} is a symmetric, general-purpose Hessian ensemble at 68\% CL obtained through a META/PDF4LHC21\_40 combination~\cite{Gao:2013bia,PDF4LHCWorkingGroup:2022cjn} of eight dynamic-tolerance ensembles. Among the alternative solutions, it includes the negative small-$x$ gluon behavior explored in CT18X. It therefore constructs a Hessian representation incorporating both the usual statistical/aleatoric uncertainty and variations associated with an enlarged model space. The eight input ensembles are selected through exploration and combination of all statistically acceptable Cloud solutions according to the procedure summarized in Sec.~\ref{sec:ReplicaSelection}. The resulting CT25m ensemble represents PDF uncertainty through exploration of a broad space of solutions with the CT25 data baseline and diverse functional forms, producing larger error bands than CT25 NNLO.
    \item \textbf{CT25pd NNLO} is a special proton-deuteron Hessian ensemble that excludes data on heavy-nuclear targets (e.g., Cu [E605] and Fe [NuTeV, CCFR]) and applies a fixed CJ-like deuteron correction~\cite{Li:2023yda} to 
    deuteron structure function data using the numerical procedure in Ref.~\cite{Accardi:2021ysh}. Otherwise, CT25pd uses the same parametrizations and uncertainty propagation as the CT25 baseline.
    \end{enumerate}
    
    Experimentalists can use the above three PDF ensembles in their single-data-set analyses according to the procedure outlined in Eq.~(\ref{fit2}) and the text below it. In this paradigm, a POI is profiled using a selected LHC data set and the $\Delta\chi^2=1$ prescription on an exhaustive surrogate PDF model $\widetilde{\bm{\pi}}$. Our goal is to offer such a model that represents a variety of good Cloud solutions while also allowing the exploration of solutions obtained under alternative assumptions, such as nonstandard choices of fitted data sets. Within the current release, we include three ``conservative prior'' ensembles for this purpose.
    \begin{enumerate}
       \setcounter{enumi}{2} 
    \item The goal of the \textbf{CT25FlatP} (Flat Prior, Proton Only) envelope ensemble is to provide an extended selection of PDF sets that fit the proton-scattering data sets in the CT25 baseline as well as CT25 NNLO itself. Its 350 members include 100 Monte Carlo error sets generated from the nominal CT25 NNLO parametrization, supplemented by 150 as-good fits to the proton-only subset of CT25 data with alternative parametrizations, including another 100 replicas from potentially negative small-$x$ gluon. The "envelope ensemble" designation means that all members are treated as equally statistically acceptable, so the uncertainty is given by their envelope. CT25FlatP deliberately suppresses information from deuteron data. It therefore provides a proton-only prior for reweighting with LHC data, designed for predictions in which one wishes to avoid potential deuteron-model biases. Its envelope represents both aleatoric and epistemic uncertainties.
    \item To address the concern that global PDFs may absorb biases from unknown SM dynamics or new physics, we provide two ensembles with reduced statistical weights for classes of data. Comparing their profiling with that of the standard ensembles may reveal such biases. In the \textbf{CT25hh} (hadron-hadron collider) Hessian (DT) ensemble, experiments performed outside the Tevatron and LHC receive a $\chi^2$ weight of 0.1. Conversely, \textbf{CT25nhh} applies this weight to all Tevatron and LHC data sets. These choices shift the central PDFs and moderately increase their uncertainties. 
\end{enumerate}

The near-convex behavior of the total $\chi^2$ supports combining multiple parametrizations and using the approximate Gaussian description underlying the Hessian ensembles. Together, these approaches show how the exploration of the model space accounts for distinct sources of uncertainty relevant to different phenomenological applications.

\vspace{-6pt}\subsection{Selecting representative solutions from the CT25 Parton Cloud \label{sec:ReplicaSelection}}

We briefly describe how representative central fits for CT25m and CT25FlatP are selected from the CTEQ-TEA Parton Cloud. The process has three steps: characterizing and clustering the solutions, filtering those with large $\chi^2$, and bundling the remainder into a final ensemble.

In the first step, shape-based clustering groups the solutions according to their PDF shapes through Uniform Manifold Approximation and Projection (UMAP)  dimensionality reduction. At this stage, solutions are characterized by numerical vectors that reflect their shapes in $x$. This is achieved either by providing the PDF values at selected $x$ points or, alternatively, by using an absolute shape estimator based on an information criterion~\cite{Courtoy:2025ppd}. The purpose is to identify neighborhoods of qualitatively similar solutions without relying solely on their $\chi^2$. Both characterizations lead to similar clustering patterns.

The second step consists of filtering or pruning each identified cluster. Within each shape cluster, the corresponding solutions define a local region of the 
landscape. A representative solution is selected according to its goodness of fit. This distinction is important: shape determines the neighborhood, while fit quality determines which point represents it. In this way, we choose eight input fits for the general-purpose CT25m ensemble. If pruning is pursued instead, redundant solutions are removed with a complementary algorithm, yielding sets such as the special-purpose CT25FlatP.

The final step bundles the selected solutions.

\vspace{-6pt}\subsection{Flavor structure at large $x$ with four PDF ensembles}

\begin{figure}[tb]
    \centering
    \includegraphics[width=0.97\linewidth]{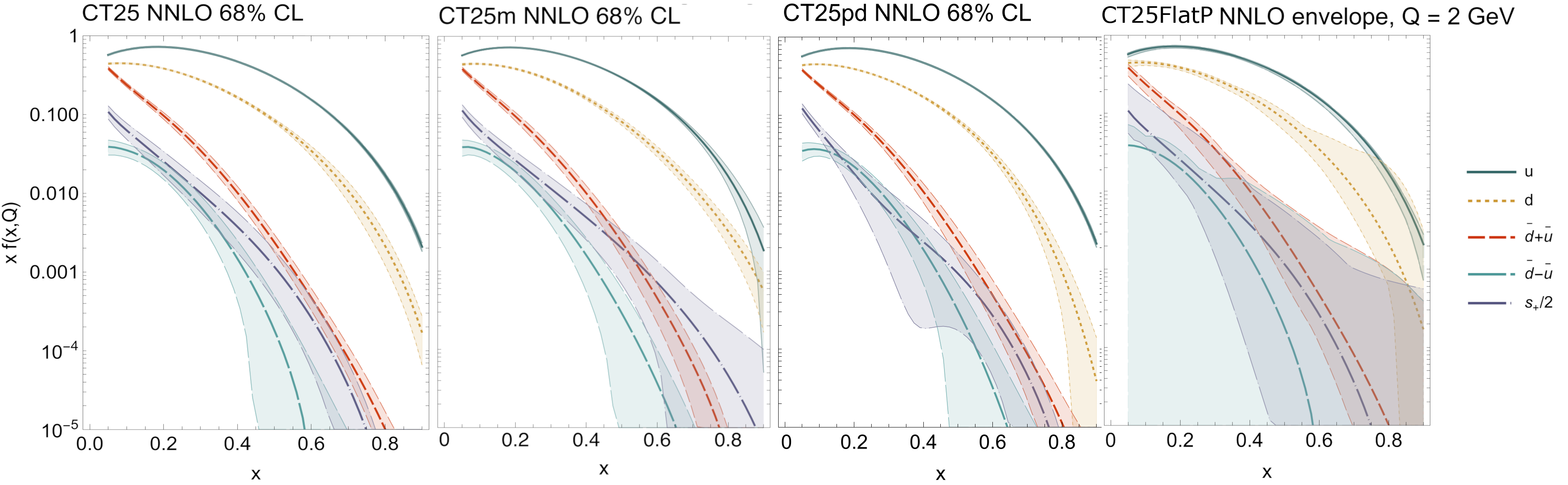}
    \caption{Large-$x$ behavior of the light sea quarks in four ensembles of the CT25 family at $Q=2$ GeV.}
    \label{fig:largeX}
\end{figure}

At $x>0.1$, relevant to BSM searches at the HL-LHC and to EIC physics, the special CT25 ensembles represent sea-quark and gluon PDFs more informatively than general-purpose PDFs. Deuterium and nuclear fixed-target data constrain flavor structure but may introduce biases relative to proton-only data. Figure~\ref{fig:largeX} compares the (anti)quark uncertainty bands in the deuteron-free CT25FlatP ensemble with those from CT25, CT25m, and CT25pd. The differences reflect data-set choices, nuclear or higher-twist contributions, and relaxed priors in unconstrained regions.

CT25m is compatible with CT25 and MSHT20 over most of the region, although its central values and uncertainties differ more at extreme large $x$, particularly for sea quarks and the gluon. CT25pd applies the CJ--CT correction for nucleon binding in deuterium, affecting the large-$x$ ratios $d/u$, $\bar d/\bar u$, and $(s+\bar s)/(\bar u+\bar d)$. The broader CT25FlatP envelope covers the CT25m 68\% CL bands and reflects a wider range of epistemic uncertainty. Similar trends appear in the small-$x$ extrapolation region.

We also see in Fig.~\ref{fig:largeX} that the $u$ and, to a lesser degree, $d$ PDFs are well-constrained, in contrast to $\bar u$, $\bar d$, and $s$, which become uncertain at $x>0.3$. Eliminating the NuTeV dimuon data in CT25pd increases $s(x)$ and enlarges its uncertainty relative to CT25 with the same parametrization. The large-$x$ strangeness uncertainty is also increased in CT25m because of its wider variety of parametrizations. Removing deuteron data in CT25FlatP also inflates the uncertainty on the $d$ quark. While CT25 NNLO realizes a "low-sea" scenario with rapidly decreasing $\bar q$ at $x>0.3$, the much wider proton-only uncertainties of CT25FlatP also allow higher $\bar q$ values at asymptotic $x$, similarly to NNPDF4.0, and even $SU(3)_\textrm{flavor}$-symmetric asymptotic behavior at the largest $x$.

The four ensembles therefore allow data and prior constraints to be removed one at a time in a controlled way to examine the sensitivity of the resulting predictions. Our future plan is to expand these studies by adding Parton Cloud fits with other choices of kinematic cuts and nuclear and low-$Q$ power-suppressed corrections, all of which affect the large-$x$ asymptotics of the PDFs in a correlated manner \cite{Courtoy:2020fex}.

\begin{figure}[t]
    \centering
\includegraphics[width=0.49\linewidth]{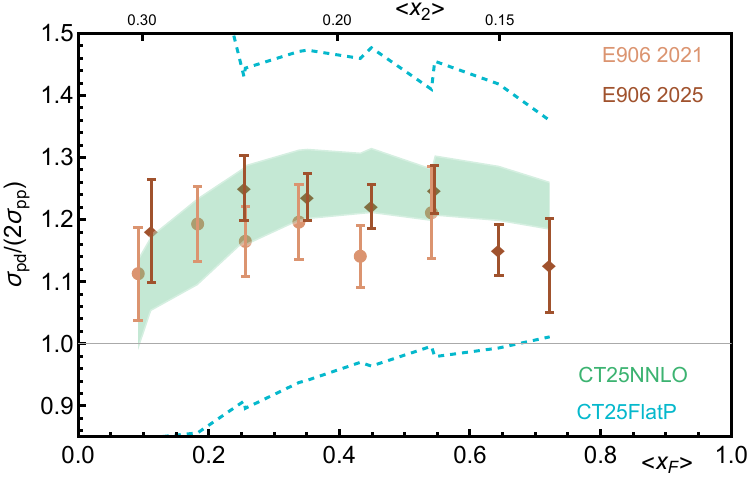}
\includegraphics[width=0.49\linewidth]{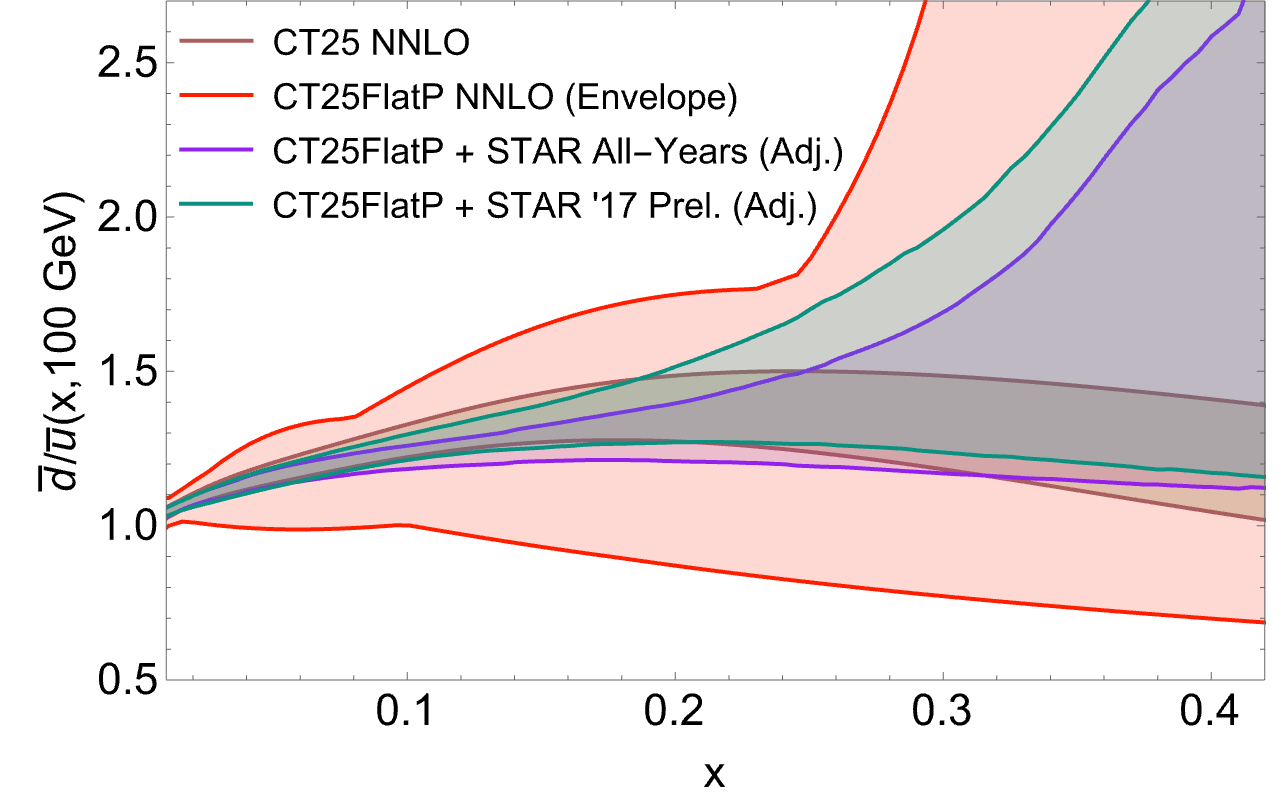}
\caption{
Left: Postdiction of the newest SeaQuest E906 (2025) data~\cite{FNALE906SeaQuest:2025kjo} and comparison with the 2021 data for CT25 NNLO (light green band) and CT25FlatP (cyan dashed contour). Right: The $\bar d(x,Q)/\bar u(x,Q)$ ratio from the deuteron-free CT25FlatP "flat prior" ensemble, reweighted by the STAR $W^+/W^-$ production ratio at $\sqrt{s}=510$ GeV~\cite{Ponce-Chavez:2026lqh}.}
\label{fig:seaquest2025}
\end{figure}

For large-$x$ PDFs, the SeaQuest E906 (2025) Drell--Yan data and the $W^+/W^-$ cross-section ratio in $pp$ collisions at STAR provide complementary information on $\bar d/\bar u$. Although not included in CT25, the E906 behavior at $x>0.1$ depends on the assumed $\bar d+\bar u$ distribution and therefore varies among global analyses, as shown for four CT25-family sets in the left panel of Fig.~\ref{fig:seaquest2025}. Because the CT25FlatP proton-only prior is unconstrained by E906 (2021), reweighting provides an unbiased test and gives an excellent postdiction of the 2021 and 2025 results. Reweighting its a priori unconstrained $\bar d/\bar u$ ratio with the STAR $W^+/W^-$ ratio gives $\bar d(x,M_W)/\bar u(x,M_W)\approx1.2$ at $\langle x\rangle=0.16$ with high precision, independently of deuteron constraints \cite{Ponce-Chavez:2026lqh}. Thus, the moderate-$x$ enhancement $\bar d/\bar u>1$ is a feature of proton PDFs rather than an artifact of deuteron dynamics, as illustrated by the narrow reweighted band in the right panel.

\vspace{-6pt}\section{Interdependence of the top-quark mass, QCD coupling, and gluon PDF}

As a phenomenological application of the CT25 analysis, we discuss a simultaneous determination of the pole top-quark mass $m_\textrm{top}^\textrm{pole}$, QCD coupling $\alpha_s(M_Z)$, and gluon PDF $g(x,Q)$ at large momentum fractions. A correlated combination of these quantities is tightly constrained by LHC data sets on the production of jets and top-quark pairs. These scattering processes are dominated by gluon initial states, while $\alpha_s(M_Z)$ and $m_\textrm{top}^\textrm{pole}$ control the magnitude of the hard-scattering cross sections. As a result, the extracted values of the three quantities affect one another.

\begin{figure}[b]
    \centering
    \includegraphics[width=0.48\linewidth]{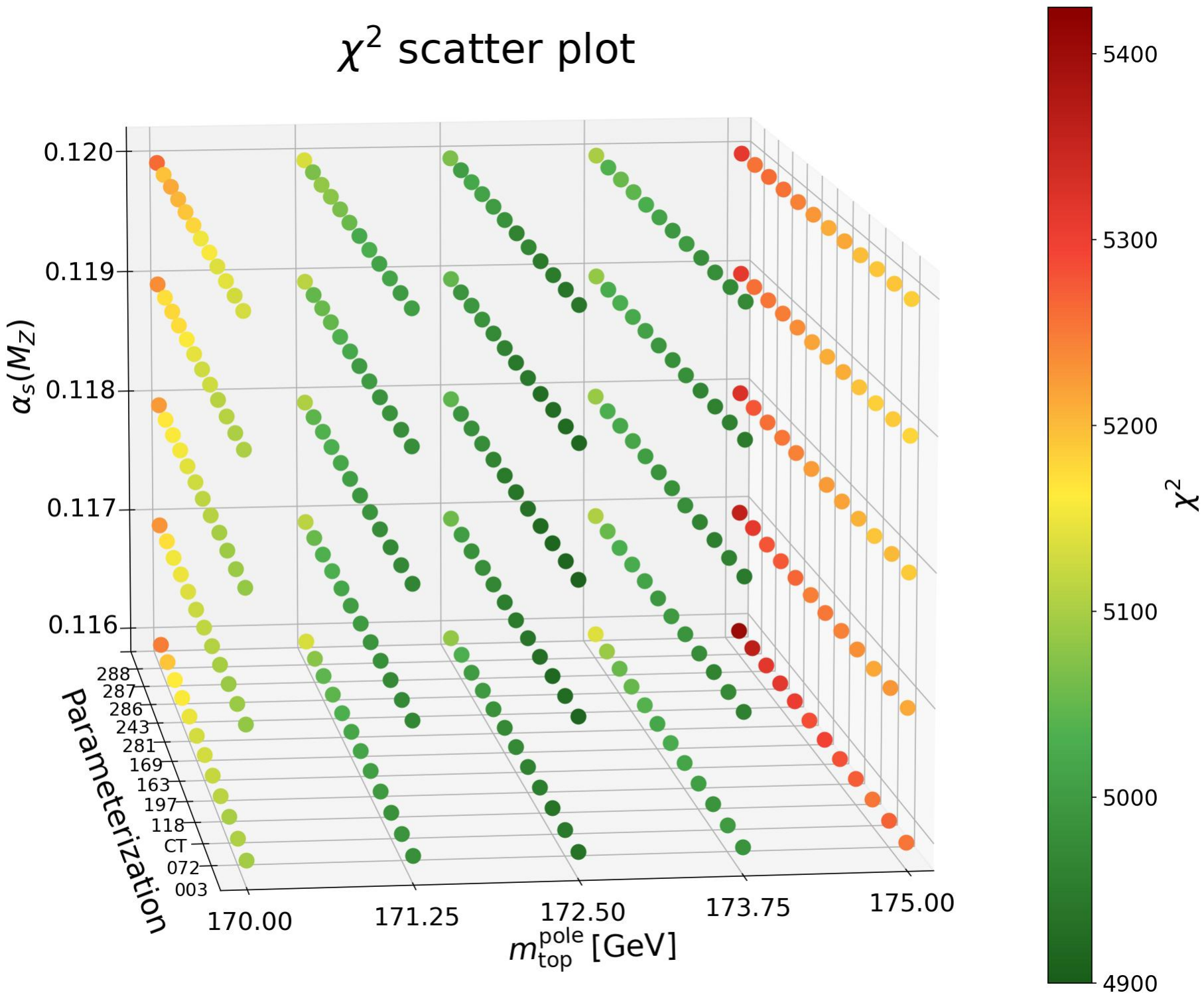}\quad
    \includegraphics[width=0.48\linewidth]{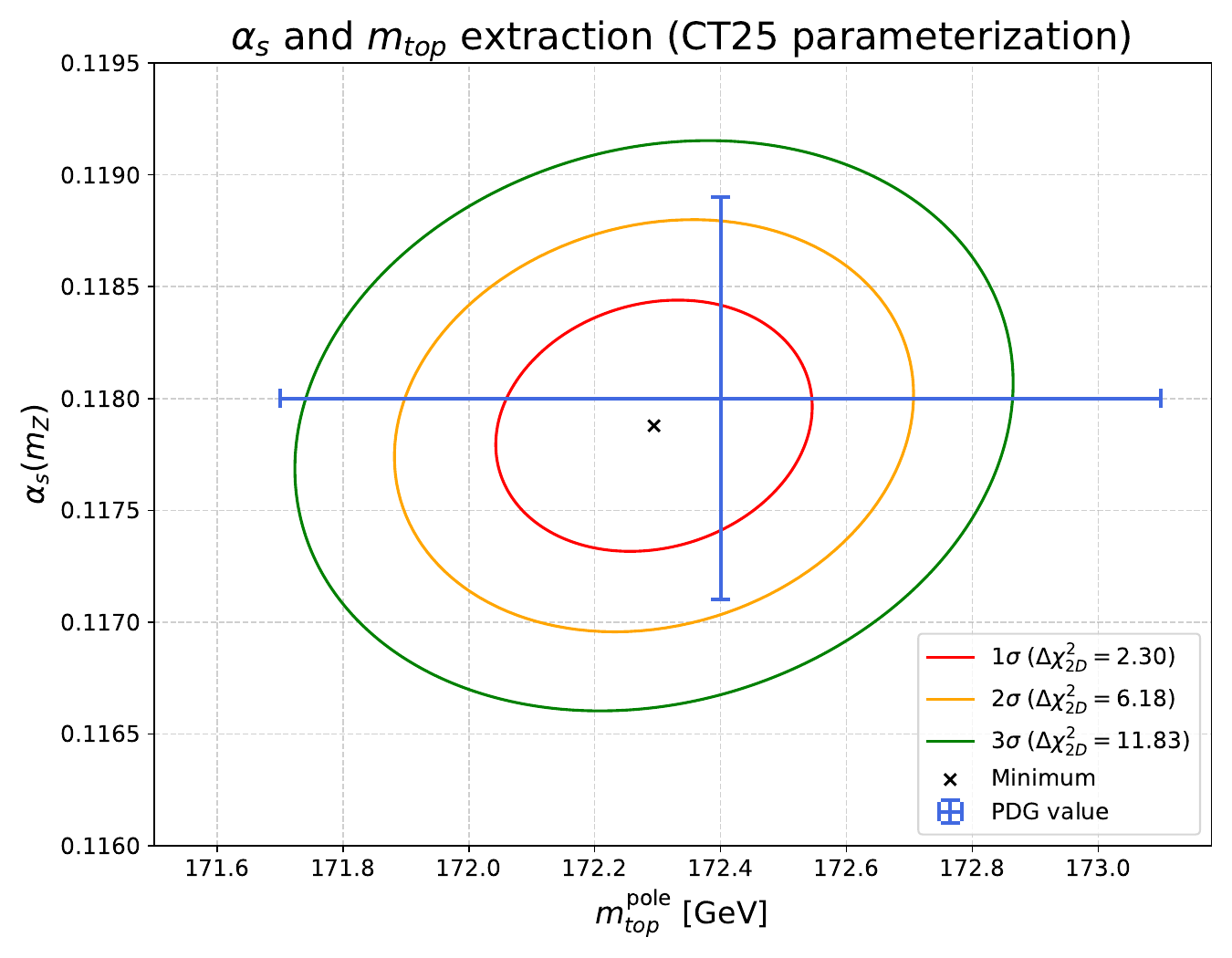}
    \caption{Left: $\chi^2$ (shown by color) for the fits using select $\alpha_s(M_Z)$, $m_\textrm{top}^\textrm{pole}$, and PDF parametrizations. Right: Best-fit values for $\alpha_s$ and $m_\textrm{top}^\textrm{pole}$ extracted from fits using the default CT25 parametrization, together with error ellipses corresponding to various confidence intervals. The PDG world average values are also shown for comparison.}
    \label{fig:mtop_3d_chi2_plot}
\end{figure}

To examine the interplay among these three components, we repeated CT25 fits on a grid of five values of $\alpha_s(M_Z)$ from $0.116$ to $0.120$, five values of $m_\textrm{top}^\textrm{pole}$ from $170$ to $175$ GeV, and twelve distinct PDF parametrizations: default CT25 NNLO and eleven representative parametrizations from the Parton Cloud. We examined the goodness-of-fit $\chi^2$ for each fit, with the results summarized in the left panel of Fig.~\ref{fig:mtop_3d_chi2_plot}. The value of $m_\textrm{top}^\textrm{pole}$ affects fit quality more than $\alpha_s(M_Z)$. Individual parametrizations do not uniquely prefer particular values of $\alpha_s(M_Z)$ and $m_\textrm{top}^\textrm{pole}$; they tend to improve or worsen fit quality for all combinations together.

We then determine confidence ellipses for $\alpha_s(M_Z)$ and $m_\textrm{top}^\textrm{pole}$ with the CT25 NNLO parametrization by profiling $\chi^2$ in the two-dimensional parameter space. 
The ellipses use the confidence intervals $1\sigma$, $2\sigma$ and $3\sigma$ corresponding to $\Delta\chi^2_{\mathrm{2D}}=2.30$, $6.18$, and $11.83$. The right panel of Fig.~\ref{fig:mtop_3d_chi2_plot} gives the optimal values $\alpha_s(M_Z)=0.1179\pm0.0006$ and $m_\textrm{top}^\textrm{pole}=172.29\pm0.25$ GeV. Our values appear to be fully compatible with the Particle Data Group averages~\cite{ParticleDataGroup:2024cfk}.

\vspace{-6pt}\section{Conclusions}

The CT25 analysis advances the CTEQ--TEA program through updated experimental input and in-depth exploration of PDF uncertainties.
The nominal CT25 NNLO PDFs remain compatible with CT18 within
uncertainties, while the new data sharpen constraints and modify the gluon and flavor structure in regions relevant to precision phenomenology. At the same time, tensions among data sets and the
growing role of correlated systematic errors require uncertainty estimates to be supported by diagnostics of fit quality, residuals, and nuisance parameters.

The CT25 Parton Cloud, a new repository of high-quality fits, facilitates the exploration of parametrizations, data selections, and methodological assumptions. Shape-based clustering, goodness-of-fit selection, and ensemble combination 
propagate the acquired information into final uncertainties.
CT25m incorporates eight representative fits into a general-purpose Hessian ensemble, while CT25pd, CT25FlatP, and
CT25hh/nhh probe the dependence on nuclear corrections, proton-only constraints, and relative data weights. Their differences are particularly informative at large $x$, where limited flavor separation and nonperturbative effects leave substantial freedom in the sea-quark distributions.

The strong-coupling determination, $\alpha_s(M_Z)=0.1183^{+0.0023}_{-0.0020}$ at 68\% CL,
illustrates the importance of accounting for data tensions and methodological uncertainty beyond the nominal $\Delta\chi^2=1$ prescription. The joint scans of $\alpha_s(M_Z)$, top-quark pole mass, and PDF parametrizations further demonstrate the need to examine QCD parameters together with the PDFs.
The smaller uncertainties obtained in the fixed-parametrization,
$T^2=1$ scan should therefore be distinguished from uncertainties 
that incorporate the broader variations explored in the global analysis.

Together, these developments provide a flexible framework for precision LHC predictions and for assessing PDF uncertainties in the kinematic regions relevant to the HL-LHC and EIC. Additional
flavor-sensitive measurements and improved control of theoretical and nuclear corrections will be essential to resolve the remaining ambiguities, particularly in the large-$x$ sea.

\vspace{-6pt}\section*{Acknowledgments} 
The authors acknowledge the use of the OpenAI's ChatGPT 5.6 to prepare an initial outline of their presentations and are solely responsible for the content, interpretations, and conclusions. 

AC is supported by UNAM Grant No. DGAPA-PAPIIT  IN102225 and the Secihti grant CBF-2026-49; part of her contribution to this work was performed at the Aspen Center for Physics, which is supported by a grant from the Simons Foundation (1161654, Troyer). 
This work is supported in part by the U.S. National Science Foundation under Grants No. PHY-2310291 and PHY-2412071 and by the U.S. Department of Energy under Award DE-SC0026593 and Contract DE-AC02-06CH11357. The work at MSU is partly supported by the Wu-Ki Tung Endowed Chair in Physics and used computational resources provided by the MSU Institute for Cyber-Enabled Research.


\bibliographystyle{JHEP}
\bibliography{biblio}
\end{document}